
\footline={\ifnum\pageno=1\firstfootline\else\otherfootline\fi}
\def\firstfootline{\rm\hss\folio\hss}
\def\otherfootline{\hfil}
\font\tenbf=cmbx10
\font\tenrm=cmr10
\font\tenit=cmti10
\font\elevenbf=cmbx10 scaled\magstep 1
 1
 1
\font\twelverm=cmr10 scaled\magstep 1

\overfullrule=0pt

\line{\hfill BROWN-HET-873}
\line{\hfill ANL-HEP-CP-92-71}
\line{\hfill August 1992}
\vglue 1cm
\hsize=6.0truein
\vsize=8.5truein
\parindent=3pc
\baselineskip=10pt
\centerline{{\tenbf COLOR--SEXTET QUARK CONDENSATION IN
QCD}\foot{Presented to the Workshop on QCD Vacuum Structure and its
Applications, Paris, France, 1-5 June, 1992.}}
\vglue 12pt
\centerline{\tenrm KYUNGSIK KANG}
\baselineskip=13pt
\centerline{\tenit Physics Department, Brown University, Box 1843,
}
\baselineskip=12pt
\centerline{\tenit Providence, RI 02912, USA}
\vglue 0.3cm
\centerline{\tenrm ALAN R. WHITE}
\centerline{\tenit High Energy Phsyics Division, Argonne National
Laboratory}
\centerline{\tenit Argonne, IL  60439, USA}
\vglue 0.8cm
\centerline{\tenrm ABSTRACT}
\vglue 0.3cm
{\rightskip=3pc
\leftskip=3pc
\tenrm\baselineskip=12pt
\noindent
We present the minimal sextet--quark condensation model as an
attractive alternative to the standard model. The model is constructed
by a simple and natural extension beyond the standard model and has
many new interesting features some of which can be tested readily at
the existing colliders.  A crucial test may be a short--lived axion--
like $\eta_6$ of mass 30--60 GeV, which can produce high energy
photons diffractively at hadron colliders and radiatively in $e^+ e^-$
colliders.
\vglue 0.8cm}
\line{\elevenbf 1.  Introduction\hfil}
\vglue 0.4cm
\baselineskip=14pt
\twelverm
The idea of technicolor$^1$ is to search for an alternative to spontaneous
symmetry breaking (SSB) and to break
the electroweak gauge symmetry {\bf dynamically}, without the use of
Higgs scalars, while preserving the successful mass relation of the
weak gauge bosons.  There are two possibilities to achieve dynamical
symmetry breaking (DSB):  either$^1$ one introduces a QCD--like
technicolor group (TG) which confines at $\Lambda_{TC} \sim 1 $TeV,
along with anomaly--free techniquark doublets $\psi$ in a fundamental
representation of TG but in singlets of the QCD group, or$^2$ one uses
extra heavy quarks belonging to some higher dimensional representation
of the familiar QCD group $SU(3)$.  In the first approach, the global
chiral symmetry $SU(2)_L \times SU(2)_R$ in the technicolor sector is
broken down to $SU(2)$ by techniquark condensate $\langle
\bar{\psi}_L \psi_R \rangle \sim \bigl( \Lambda_{TC}\bigr)^3$,
causing a triplet of massless composite technipions which are absorbed
by the weak gauge bosons to become their longitudinal components.  It
is then possible to obtain $M_W = M_Z \cos \theta_W = {g\over 2}
\Lambda_{TC}$.  To generate the quark and lepton masses within a
dynamical context, one needs to embed TG in a larger extended
technicolor group (ETG)$^3$ which is broken down to TG at a scale
$\Lambda_{{\rm ETC.}} > \Lambda_{TC}$ and gives mass to the ordinary
quarks and leptons radiatively through the heavy ETG gauge boson
couplings to fermions and technifermions.  This gives typically a
fermion mass $m_f\sim \Lambda_{TC}^3 /\Lambda_{{\rm ETC.}}^2$.  However the
ETG gauge boson couplings cause unwanted flavor--changing neutral
current (FCNC) interactions in the ordinary quark sector due to
fermion--fermion couplings.  The $K_L - K_S$ mass difference, a
$\Delta S = 2$ process, leads to $\Lambda_{{\rm ETC}} > $500
TeV, which in turn predicts a typical fermion mass $m_f$ to be a few
MeV at the most.  A way to overcome the FCNC difficulty is to force
the technicolor coupling constant to run very slowly$^4$, i.e., to ``walk"
by considering the technicolor theory near a fixed point of the beta
function that describes the evolution of the technicolor coupling
constant.  The assumption is that the beta functions for the
technicolor coupling is small above $\Lambda_{TC}$ scale and the
anomalous dimension of the fermion mass operator can be large all the
way up to $\Lambda_{{\rm ETC.}}$.  In such cases, the techniquark
condensate is given by $\langle \bar{\psi}_L \psi_R\rangle \sim
\Lambda_{TC}^3 \bigl( \Lambda_{{\rm ETC}} /\Lambda_{TC}
\bigr)^{\gamma}$ where $\gamma$ is the anomalous dimension of the
bilinear operator $\bar{\psi}_L \psi_R$.  Thus it is possible to
achieve less spectacular enhancements of the techniquark condensate
for a walking technicolor theory in which asymptotic freedom is
nearly saturated.  Of course care must be given to maintain the
stability of the fixed point and asymptotic
freedom even when higher--loop corrections are taken into account.

In the second approach of DSB, new quarks belonging to a higher
color representation of the ordinary color $SU(3)$ group of QCD play
the role of the techniquarks of the technicolor model, provided that
they condense at the scale required for the dynamical electroweak
symmetry breaking.  In fact it is well--known that the ordinary quarks
can form condensates at a scale $\leq 1$GeV, which then break the
global chiral symmetry and produce pions as the Goldstone bosons.
This scale is obviously far too small to give the right mass relation
for the weak gauge bosons.  Recently several authors$^5$ suggested
heavy top quark condensation as a possibility of DSB because of
the experimental lower bound $m_t > 89$ GeV.  Detailed investigation of
the minimal top quark condensation model with the four fermion
interaction $\bigl( \bar{\psi}_L t_R \bigr) \bigl( \bar{t}_R
\psi_L\bigr)$, $\psi_L$ being the third generation doublet $(t_L ,
b_L)$, predicts the top quark mass $m_t = 220 \sim 230$ GeV at the
scale of new physics $\Lambda \simeq 10^{15} \sim 10^{19}$ GeV.  The
result is dangerously close to the indirect experimental upper bound
of $m_t$ based on the analysis of radiative corrections to
the $\rho$--parameter.

With the extra quarks belonging to higher representations $R>3$ under
color $SU(3)$, it is possible to achieve much larger enhancement
of the condensate because the strength of the quark--antiquark binding
potential is proportional to $C_2 (R) \alpha_S (\mu_R )$ where $C_2 (R)$
is the quadratic Casimir invariant of the representation ${\bf R}$ and is
bigger than $C_2 (3) = 4/3$ for $R>3$.  In that case, the
fermion mass operator in exotic quark sectors can condense at a scale
much larger than 1 GeV and perhaps as large as the desired electroweak symmetry
breaking scale.  This type of DSB has a number of attractive features.
First of all,
the chiral breaking scale of QCD provides the electroweak symmetry
breaking scale, which can lead to many new phenomena due to mixing of
electroweak and strong interactions.  Secondly, it is possible to
construct$^2$ minimal DSB model with a flavor doublet of sextet quarks
that has a fixed point in the two--loop beta function and yet
preserves asymptotic freedom.  The zero seems
to survive to very high--order loop corrections and therefore the
minimal sextet--quark condensation model (MSQCM) has the required
property that one assumes for a walking technicolor model.  In
addition MSQCM can be constructed by extending naturally the gauge and
fermion sectors of the standard model under color $SU(3)$ and
electroweak $SU(2) \times U(1)$ by adding a flavor doublet of sextet
quarks.  While the model is one of the simplest extensions of the
standard model, the spectrum of the new particles is rather
unambiguous and can be tested in part at the existing colliders.  We
initially$^6$
suggested, based on the total cross--sections along with the forward
elastic $p\bar{p}$ scattering amplitude at CERN and Fermilab Tevatron
collider energies, a possible signal of color--sextet symmetry
breaking, i.e., diffractive production of an axion--like particle
$\eta_6$ originating from the sextet--quark
condensations.  In particular we pointed out that mini--Centauro and
Geminion events in very-high--energy cosmic ray experiments$^7$ could
actually be identified as the hadronic and two--photon decay products
of the $\eta_6$ respectively and led us to suggest to look for the
two--photon decay mode of $\eta_6$ with mass 30$\pm$ 10 GeV in
hadron diffractive process.  It now appears$^8$ that the most distinctive
of the Geminion events lead to a higher mass of 60-70 GeV for the
$\eta_6$.  Also it was realized that a massive $\eta_6$ would
eventually be observable as a rare radiative decay of $Z^0$ at LEP.
We emphasize that the existence of $\eta_6$ is a uniquely distinctive
sign of sextet--quark condensation in that there is no comparable
state in other models.  While there are $\eta$--like states in
technicolor models$^9$, they have either much heavier mass because of the
technicolor gauge anomaly or carry color.

In the next section, we describe briefly the structure of the MSQCM.
Section 3 contains possible evidences for the sextet--quark
condensates.  We will see that the $\eta_6$ may provide a new source
for radiative decays of the $Z^0$.

\vglue 0.6cm
\line{\elevenbf 2.  The Minimal Sextet--Quark Condensation Model
(MSQCM)\hfil}
\vglue 0.4cm

The MSQCM is a straight--forward modification of the standard $SU(3)
\times SU(2) \times U(1)$ model by adding a flavor doublet of color
sextet quarks $Q_6 = (U,D)$ with $({\bf 6}^*, {\bf 2}, 1/3)$ to the
usual fermions of the standard model but with no elementary Higgs
scalars.  There are two additional doublets of new leptons that are
required by anomaly cancellation.  The Lagrangian is given by

$${\cal L} = {\cal L}_{{\rm gauge}} + {\cal L}_{{\rm fermion}} + {\cal
L}_{4f}\eqno(1)$$
where ${\cal L}_{{\rm gauge}}$ is the Lagrangian for the color $SU(3)$
and electroweak $SU(2)$ and $U(1)$ gauge fields, ${\cal L}_{{\rm
fermion}}$ is the kinetic terms of the ordinary and additional
fermions constructed by making use of the appropriate $SU(3) \times
SU(2) \times U(1)$ covariant derivatives, and ${\cal L}_{4f}$
represents all relevant four--fermion interaction terms of scale
dimension 6 that are flavor--chiral and gauge invariant.

With the usual six flavors of ordinary quarks and two flavors of
sextet quarks, asymptotic freedom is still preserved in MSQCM but
there is a zero in the QCD beta function

$$\mu {\partial g\over \partial\mu} = - {g^3\over 16\pi^2} \left[ b_0
+ b_1 \left( {g^2\over 16\pi^2}\right) + b_2 \left( {g^2\over
16\pi^2}\right)^2 + \cdots \right] = \beta (g)\eqno(2)$$
where
$$\eqalign{ b_0 = & {11\over 3} C(A) - {4\over 3} \sum_R I_R N_R = 11
- {2\over 3} (N_3 + 5N_6 ) = {1\over 3}\cr
b_1 = & + {34\over 3} C (A)^2 - {20\over 3} C(A) \sum_R I_R N_R - 4
\sum_R C(R) I_R N_R\cr
= & 102 - {38\over 3} N_3 - {250\over 3} N_6 = - 140 {2\over
3}}\eqno(3)$$

\noindent
and $b_2$ is given$^{10}$, in the momentum subtraction scheme, approximately
by $b_2 \simeq 0.01 (b_1 /b_0 )^2$.  Here $C(A)$ and $C(R)$ are the
quadratic Casimir operators of the adjoint and ${\bf R}$--
representations respectively and $I_R$ and $N_R$ denote the second
index and multiplicity of the representation ${\bf R}$.  Notice that
the stability of the fixed point is guaranteed if $b_0 b_2 /b_1^2 <
{1\over 4}$, which is well satisfied by the MSQCM.  From the smallness of
the third term in Eq. (2) for $\alpha_S = {g^2\over 4\pi} = - 4\pi
b_1 /b_0 \simeq {1\over 33}$ and of the expansion parameter
$g^2 /16\pi^2 = \alpha_S /4\pi \simeq 1/412 $ it is clear that the
zero should survive to
very high order in the perturbation expansion of $\beta (g)$.  In
other words, the MSQCM has the desired property of the walking
technicolor.

There is a $U(2) \times U(2)$ chiral flavor symmetry of the
sextet--color quark sector in MSQCM.  Normal QCD chiral dynamics may
cause chiral--symmetry breakdown in the sextet quark sector at a mass
scale $F_{\pi_6} \sim 250$ GeV, much higher than the usual scale
$F_{\pi}$ of the chiral symmetry breaking in the color--triplet quark
sector.  This can be understood from an elementary picture in which
the strength of the effective quark--antiquark binding potential is
propotional to $C_2 (R) \alpha_S (\mu_R )$ and chiral symmetry
breaking occurs at some common critical value of the flux string
tension so that $C_2 ({\bf 6} ) \alpha_S (F_{\pi_6} )\sim C_2 ({\bf 3}
) \alpha_S (F_{\pi} )$.  As the QCD coupling $\alpha_S (\mu_6 )$
evolves (actually ``walks" slowly), a $\bar{Q}_6 Q_6$ condensates
forms at a scale $F_{\pi_6} = \vert\langle \bar{Q}_6 Q_6 \rangle
\vert^{1/3} \sim 250$ GeV triggering the chiral--symmetry break--down
spontaneously and producing four massless Goldstone bosons, an
isotriplet ${\bf \pi}_6$ and an isosinglet $\eta_6$. Then through the coupling
terms of the weak gauge boson fields of the standard model to the
sextet--quark currents in ${\cal L}_{{\rm fermion}}$, the isotriplet
${\bf \pi}_6$ are absorbed by the gauge bosons to become the
longitudinal components of $W^{\pm}$ and $Z^0$ with the mass relation
$M_W = M_Z \cos \theta_w \sim gF_{\pi_6}$.  In analogy with the
familiar color--triplet quark sector, ${\bf \pi}_6$ and $\eta_6$
couple longitudinally to the isotriplet and
isosinglet (under the unbroken $SU(2)$) axial--vector currents
respectively.  Actually the isosinglet axial--vector current would
contain a small admixture$^{11}$ of color--triplet quarks so that it
is free of the QCD $U(1)$--anomaly and transforms orthogonal under the
$U(1)$ flavor symmetry to another combination which is made mostly
of the isosinglet axial--vector current of the color--triplet quarks
with a small color--sextet component and couples to the QCD $U(1)$--
anomaly.  We note that $\eta_6$ is effectively an axion since it is
the Goldstone boson associated with a $U(1)$ axial chiral symmetry of
the color--sextet quarks.  As pointed out above, in the MSQCM,
$\alpha_S$ evolves very slowly above the electroweak scale
and interactions involving small instantons,
integrated over a large momentum range, can produce$^{12}$ a large anomalous
dimension for sextet quark operators and in particular a much larger
$\eta_6$ mass than a conventional axion mass of ${\cal O}$(100keV).

The existence of the $\eta_6$ provides a unique signal of sextet quark
symmetry breaking, which is unparalleled by other models, because one
generally does not expect any new strong interaction of the Higgs
sector, elementary or dynamical, and the technicolor sector
to manifest themselves until way above
the electroweak scale.  The $\eta_6$ of the MSQCM allows the early
appearance of such a new interaction.  In elementary Higgs models,
the analogous Higgs scalar to the $\eta_6$ do not give either the
anomalous two--photon decay or the off--shell non--local coupling that
can allow an additional source for photon production.  Of course, there
will be in MSQCM new QCD baryons and mesons of pseudoscalar, vector,
axial--vector etc. states that are made of sextet quarks, but they are
all expected to be heavy, as the color--sextet quark masses are expected to
be larger than $F_{\pi_6}$.  In fact, an estimate based on
the nontrivial solution of the ladder Schwinger--Dyson equation for the
$Q_6$ propagator in the quenched planar approximation gives$^{13}$ the
color--sextet quark mass to be about 350 GeV for $t$--quark mass
range of 77--160 GeV.  In short the existence of a
particle with the properties of the $\eta_6$ is a very special
signal of the existence of a color--sextet quark sector that
can be tested at the existing collider energies.
\vglue 0.6cm
\line{\elevenbf 3.  $\eta_6$ Production and Decay\hfil}
\vglue 0.4cm

Because of the interplay between QCD and electroweak interactions in
MSQCM, a host of new phenomena is expected$^6$ at the energy scale $>
F_{\pi_6}$.  In addition, there will be new baryons and mesons which
are made of color--sextet quarks.  They are all expected to be heavy,
i.e., $M\geq M_{Q_6}$, some of which may already have been hinted by
the high--energy cosmic exotics.  The cross--section for perturbative
QCD production$^{15}$ of $Q_6$ is, apart from color factors, the same as that
of color--triplet quarks having the same mass $M_{Q_6}$ and the hard
cross--section is expected to be smaller than 1 Pb at the Fermilab
Tevatron collider.  Other possibly related new phenomena include the
existence$^{14}$ of higher order composite operator contributions to hadron
final states with at least six jets which may be observable at LHC and
strong production$^{15}$ of $W^+ W^-$ and $Z^0 Z^0$ pairs which also may have
been indicated already by experiments.
However the effects of the $\eta_6$, as an axion--like particle
associated with the axial $U(1)$ symmetry of color--sextet quarks, may
be much easier to detect$^{16}$.  We initially proposed$^6$ to look for a
massiv
$\eta_6$ via its two photon decay made in hadronic diffractive
processes.  We argued that the recent measurements of high energy
elastic $p\bar{p}$ scattering amplitude and total cross sections at
CERN and Fermilab collider energies might be due to diffractive
production of the $\eta_6$ that has the properties of two particular
classes of cosmic events known as ``geminions" or ``binocular" events and
``mini--centauros".  Diffractive production of the $\eta_6$ may be an
explanation of the puzzle of small diffractive cross sections
at collider energies.

As a light axion, the $\eta_6$ can have a major two--photon decay
modes via the anomaly.  Using the usual PCAC relation between the
isosinglet axial--vector current and the $\eta_6$ field, the decay
rate of the $\eta_6$ into two photons can be calculated from anomaly,
giving

$$\Gamma (\eta_6 \rightarrow 2\gamma ) \sim \left( m_{\eta_6}^3
\alpha_{em}^2 e_q^4 / 16\pi^3 F_{\eta_6}^2 \right) N^2 \sim \left(
{100 keV\over m_{\eta_6}} \right)^2 {\rm sec}\eqno(4)$$

\noindent where $e_Q$ is the charge of sextet quarks,  $N=6$ is the
color number, and $F_{\eta_6}$ is the decay constant.  If $m_{\eta_6}
\sim 60$ GeV, the lifetime of the $\eta_6$ from Eq. (4) is comparable
with that of the $\pi^0$ for $F_{\eta_6} \sim  F_{\pi_6} \sim 250$GeV,
i.e., $3\times 10^{-17}$sec. The $\eta_6$ can be produced$^{17}$ radiatively
from $Z^0$.  The branching ratio for $Z^0 \rightarrow \eta_6 + \gamma$
is calculable from the triangle anomaly giving $\Gamma \left( Z^0
\rightarrow \eta_6\gamma\right) / \Gamma \left( Z^0 \rightarrow \mu^+
\mu^-\right) \sim 10^{-5}$.  This decay should eventually be
observable at LEP.  In this connection, it is interesting to speculate
that the two events of $Z^0\rightarrow \mu^+\mu^- \gamma\gamma$ with
$M_{2\gamma} \sim 60$GeV observed by the $L_3$
experiment$^{18}$ at LEP may be related with $\eta_6$ production.

A strongly interacting $Q_6$ sector has an absorption effect for $e^+
e^-$ pair production via the interference between the diffractive
excitation of a photon into a $\bar{Q}_6 Q_6$ state that decays into
$e^+ e^-$ and the electromagnetic production of $e^+ e^-$ pairs.  The
production of hadrons and $2\gamma$ via the $\eta_6$, together with
direct production of the $Z^0$, will drastically modify the properties
of high--energy photon--initiated air showers and the development of
electromagnetic clusters within hadron initiated air showers.  This
may be an explanation of the muon--rich photon showers$^{19}$ and wide range
of anomalous shower development seen in high--energy cosmic ray
events.

Besides the CDF group which$^{20}$ has been attempting to search
for the $\eta_6$ in the diffractive $p\bar{p}$ scattering, an
interesting physics possibility at a photon linear collider has been
proposed$^{21}$ to search for a host of new particles including the $\eta_6$
that have appreciable two--photon couplings.  In the meantime, the UA4-2
group has repeated the original $UA(4)$ experiment.  Diffractive
production of the $\eta_6$ would result in characteristically distinct
energy behaviors in total cross--section, which can readily be checked
at existing collider energies.  In any case, the existence of the
$\eta_6$ provides a new hitherto unknown source for photon production
from $Z^0$.

In short, the MSQCM has a number of attractive features.  It is by far
the most natural and the simplest extention of the standard model
without elementary Higgs scalars and has a uniquely distinguished advantage in
t
it can be tested at the existing colliders.
\vglue 0.6cm

\line{\elevenbf 4.  Acknowledgements \hfil}
\vglue 0.4cm
This work is supported in part by the U. S.
Department of Energy, Contract DE-AC02-76-ER03130.A31 (Report No.
BROWN-HET-873) and Contract W-31-109-ENG-38 (Report No.
ANL-HEP-CP-92-71).
\vglue 0.6cm

\line{\elevenbf 5.  References \hfil}
\medskip

\item{1.}  S. Weinberg, {\it Phys. Rev.} {\bf D13}, 974 (1976); {\bf
D19}, 1277 (1979); L. Susskind, {\it Phys. Rev.} {\bf D20}, 2619
(1979).
\item{2.}  W. J. Marciano, {\it Phys. Rev.} {\bf D21}, 2425 (1980);
E. Braaten, A. R. White and C. R. Willcox, {\it J. Mod. Phys.} {\bf
A1}, 693 (1986); K. Kang and A. R. White, {\it J. Mod. Phys.} {\bf
A2}, 409 (1987).
\item{3.}  S. Dimopoulos and L. Susskind, {\it Nucl. Phys.} {\bf
B155}, 237 (1979); E. Eichten and K. Lane, {\it Phys. Lett.} {\bf
B90}, 125 (1980).
\item{4.}  B. Holdom, {\it Phys. Rev.} {\bf D24}, 1441 (1981).
\item{5.}  V. A. Miranski, M. Tanabastin and K. Yamawaki, {\it Mod.
Phys. Lett.} {\bf A4}, 1043 (1989); W. J. Marciano, {\it Phys. Rev.
Lett.} {\bf 62}, 2793 (1989); W. A. Vardeen, C. T. Hill and M.
Lindner, {\it Phys. Rev.} {\bf D41}, 1647 (1990).
\item{6.}  K. Kang and A. R. White, {\it Phys. Rev.} {\bf D42}, 2425
(1989).
\item{7.}  S. Hasegawa, in VI International Symposium on Very High
Energy Cosmic Ray Interactions, Taubes, France (1990).
\item{8.}  T. Arizawa, Private Communication.
\item{9.}  K. Lane, Snowarass Proceedings (1982).
\item{10.}  O. V. Tarasov, A. A. Vladimirov and A. Yu Zharkov, {\it
Phys. Lett.} {\bf 93B}, 429 (1980); A. R. White, ANL-HEP-PR-84-31.
\item{11.}  T. E. Clark, C. N. Leung, S. T. Love and J. L.Rosner, {\it
Phys. Lett.} {\bf B177}, 413 (1986).
\item{12.}  B. Holdom and M. E. Peskin, {\it Nucl. Phys.} {\bf B208},
397 (1982).
\item{13.}  K. Fukazawa, T. Muta, J. Saito, I. Watanabe and M.
Yonegawa, {\it Prog. Theor. Phys.} {\bf 85}, 111 (1991).
\item{14.}  R. E.Chivukula, M. Golden and E. H. Simmons, {\it Phys.
Lett.} {\bf B207}, 453 (1991); E. H. Simmons, {\it Phys. Lett.} {\bf
B259}, 125 (1991).
\item{15.}  A. R. White, {\it Mod. Phys. Lett.} {\bf A2}, 397 (1987).
\item{16.}  K. Kang and A. R. White,  Proc. Beyond the Standard
Model II, Univ. of Oklahoma, OK (1990); Proc. Joint International
Lepton-Photon Symp. \& Europhysics Conf. on H.E. Phys., Geneva,
Switzerland (1991).
\item{17.}  T. Hatsuda and M. Umezawa, {\it Phys. Lett.} {\bf B254},
493 (1991).
\item{18.}  J. Wenninger - L3 Collaboration ,in Rencontre de Moriond (1992).
\item{19.}  G. Yodh, {\it Nucl. Phys. B} (Proc. Suppl) {\bf 12}, 277
(1990).
\item{20.}  N. D. Giokaris et al, {\it Nucl. Phys. B} (Proc. Suppl.)
{\bf 25B}, 40 (1992).
\item{21.}  D. L. Borden, D. Bauer and D. O. Caldwell, SLAC-Pub-5715.
\end